\documentclass[journal]{IEEEtran}
\usepackage{amssymb}
\usepackage{amsmath}
\usepackage{bbm}
\usepackage{array}
\usepackage{algorithm}
\usepackage{algorithmic}
\usepackage[super]{nth}
\usepackage{graphicx}
\usepackage{enumerate}
\usepackage{url}

\newcommand{\cS}{\mathcal{S}}
\newcommand{\cA}{\mathcal{A}}
\newcommand{\cR}{\mathcal{R}}

\newcommand{\be}{\begin{equation}}
\newcommand{\ee}{\end{equation}}

\newcommand{\Exp}{\mathbb{E}}

\newcommand{\ind}[1]{\mathbbm{1}_{\{#1\}}}   

\begin{document}

\title{Reinforcement Learning-based Resource Allocation in Fog RAN for IoT with Heterogeneous Latency Requirements}
\author{\IEEEauthorblockN{Almuthanna Nassar, and Yasin Yilmaz,~\IEEEmembership{Member,~IEEE}}\\
\IEEEauthorblockA{Electrical Engineering Department, University of South Florida, Tampa, FL 33620, USA}
\\E-mails: \{atnassar@mail.usf.edu; yasiny@usf.edu\}
}

\maketitle


\begin{abstract}
In light of the quick proliferation of Internet of things (IoT) devices and applications, fog radio access network (Fog-RAN) has been recently proposed for fifth generation (5G) wireless communications to assure the requirements of ultra-reliable low-latency communication (URLLC) for the IoT applications which cannot accommodate large delays. Hence, fog nodes (FNs) are equipped with computing, signal processing and storage capabilities to extend the inherent operations and services of the cloud to the edge. We consider the problem of sequentially allocating the FN's limited resources to the IoT applications of heterogeneous latency requirements. For each access request from an IoT user, the FN needs to decide whether to serve it locally utilizing its own resources or to refer it to the cloud to conserve its valuable resources for future users of potentially higher utility to the system (i.e., lower latency requirement). We formulate the Fog-RAN resource allocation problem in the form of a Markov decision process (MDP), and employ several reinforcement learning (RL) methods, namely Q-learning, SARSA, Expected SARSA, and Monte Carlo, for solving the MDP problem by learning the optimum decision-making policies. We verify the performance and adaptivity of the RL methods and compare it with the performance of a fixed-threshold-based algorithm. Extensive simulation results considering 19 IoT environments of heterogeneous latency requirements corroborate that RL methods always achieve the best possible performance regardless of the IoT environment.
\end{abstract}

\begin{IEEEkeywords}
Resource Allocation, Fog RAN, 5G Cellular Networks, Low-Latency Communications, IoT, Markov Decision Process, Reinforcement Learning.
\end{IEEEkeywords}
\section{Introduction}
\label{introduction}
There is an ever-growing demand for wireless communication technologies due to several reasons such as the increasing popularity of Internet of Things (IoT) devices, the widespread use of social networking platforms, the proliferation of mobile applications, and the current lifestyle that has become highly dependent on technology in all aspects. It is expected that the number of connected devices worldwide will reach three times the global population in 2021 with 3.5 devices per capita. However, in some regions, such as North America, the number of connected devices is projected to reach about 13 devices per capita by 2021, which makes the massive IoT a very common concept. This trend of massive IoT will generate an annual global IP traffic of 3.3 zettabytes by 2021, which corresponds to 3-times the traffic in 2016 and 127-times the traffic in 2005, in which wireless and mobile devices will account for the 63\% of this forecast \cite{r1}. This unprecedented demand for mobile data services makes it unbearable for service providers with the current third generation (3G) and fourth generation (4G) networks to keep pace with it \cite{r2}.
The design criteria for fifth generation (5G) wireless communication systems will include providing ultra-low latency, wider coverage, reduced energy usage, increased spectral efficiency, more connected devices, improved availability, and very high data rates of multi giga-bit-per-second (Gbps) everywhere in the network including cell edges \cite{r3}. Several radio frequency (RF) coverage and capacity solutions are proposed to fulfill the goals of 5G including, beamforming, carrier aggregation, higher order modulation, and dense deployment of small cells \cite{r6}. Millimeter-wave (mm-wave) frequency range is likely to be utilized in 5G because of the spacious bandwidths available in these frequencies for cellular services \cite{r4}. Massive multi-input-multi-output (MIMO) is potentially involved for excellent spectral efficiency and superior energy efficiency \cite{r5}.

To cope with the growing number of IoT devices and the increasing amount of traffic for better user satisfaction, cloud radio access network (C-RAN) architecture is suggested for 5G, in which a powerful cloud controller (CC) with pool of baseband units (BBU) and storage pool supports large number of distributed remote radio units (RRU) through high capacity fronthaul links \cite{r7,r8}. The C-RAN is characterized by being clean as it reduces energy consumption and improves the spectral efficiency due to the centralized processing and collaborative radio \cite{r9}. However, in light of the massive IoT applications and the corresponding generated traffic, C-RAN structure places a huge burden on the centralized CC and its fronthaul, which causes more delay due to limited fronthaul capacity and busy cloud servers in addition to the large transmission delays \cite{r10,r11}.

\vspace{-2mm}
\subsection{F-RAN and Heterogeneous IoT}
The latency issue in C-RAN becomes critical for IoT applications that cannot tolerate such delays. And that is the reason fog radio access network (F-RAN) is introduced for 5G, where fog nodes (FN) are not only limited to perform RF functionalities but also empowered with caching, signal processing and computing resources \cite{r12,r13}. This makes FNs capable of independently delivering network functionalities to end users at the edge without referring them to the cloud to tackle the low-latency needs.

IoT applications have various latency requirements. Some applications are more delay-sensitive than others, while some can tolerate larger delays. Hence, especially in a heterogeneous IoT environment with various latency needs, FN must allocate its limited and valuable resources in a smart way. In this work, we present a novel framework for resource allocation in F-RAN for 5G by employing reinforcement learning methods to guarantee the efficient utilization of limited FN resources while satisfying the low-latency requirements of IoT applications \cite{r14,r15,r16}.

\vspace{-2mm}
\subsection{Literature Review}
For the last several years, 5G and IoT related topics have been of great interest to many researchers in the wireless communications field. Recently, a good number of works in the literature focused on achieving low latency for IoT applications in 5G F-RAN. For instance, resource allocation based on cooperative edge computing has been studied in \cite{r17,r18,r19,r20,r21} for achieving ultra-low latency in F-RAN. The work in \cite{r17} proposed a mesh paradigm for edge computing, where the decision-making tasks are distributed among edge devices instead of utilizing the cloud server. The authors in \cite{r18,r21} considered heterogeneous F-RAN structures including, small cells and macro base stations, and provided an algorithm for selecting the F-RAN nodes to serve with proper heterogeneous resource allocation. The number of F-RAN nodes and their locations have been investigated by \cite{r35}. Content fetching is used in \cite{r7,r19} to maximize the delivery rate when the requested content is available in the cache of fog access points. In \cite{r23}, cloud predicts users' mobility patterns and determines the required resources for the requested contents by users, which are stored at cloud and small cells. The work in \cite{r20} addressed the issue of load balancing in fog computing and used fog clustering to improve user's quality of experience. The congestion problem, when resource allocation is done based on the best signal quality received by the end user, is highlighted in \cite{r24,r25}. The work in \cite{r24} provided a solution to balance the resource allocation among remote radio heads by achieving an optimal downlink sum-rate, while \cite{r25} offered an optimal solution based on reinforcement learning to balance the load among evolved nodes for the arrival of machine-type communication devices. To reduce latency, soft resource reservation mechanism is proposed in \cite{r26} for uplink scheduling. The authors of \cite{r27} presented an algorithm that works with the smooth handover scheme and suggested scheduling policies to ease the user mobility challenge and reduce the application response time. Radio resource allocation strategies to optimize spectral efficiency and energy efficiency while maintaining a low latency in F-RAN are proposed in \cite{r28}. With regard to learning for IoT, \cite{r29} provided a comprehensive study about the advantages, limitations, applications, and key results relating to machine learning, sequential learning, and reinforcement learning. Multi-agent reinforcement learning was exploited in \cite{r30} to maximize network resource utilization in heterogeneous networks by selecting the radio access technology and allocating resources for individual users. The model-free reinforcement learning approach is used in \cite{r31} to learn the optimal policy for user scheduling in heterogeneous networks to maximize the network energy efficiency. Resource allocation in non-orthogonal-multiple-access based F-RAN architecture with selective interference cancellation is investigated in \cite{r22} to maximize the spectral efficiency while considering the co-channel interference. With the help of task scheduler, resource selector, and history analyzer, \cite{r33} introduced an FN resource selection algorithm in which the selection and allocation of the best FN to execute an IoT task depends on the predicted run-time, where stored execution logs for historical performance data of FNs provide realistic estimation of it. Radio resource allocation for different network slices is exploited in \cite{r34} to support various quality-of-service (QoS) requirements and minimize the queuing delay for low latency requests, in which network is logically partitioned into a high-transmission-rate slice which supports ultra-reliable low-latency communication (URLLC) applications, and a low-latency slice for mobile broadband (MBB) applications.

\vspace{-2mm}
\subsection{Contributions}
With the motivation of satisfying the low-latency requirements of heterogeneous IoT applications through F-RAN, we provide a novel framework for allocating limited resources to users that guarantees efficient utilization of the FN's limited resources. In this work, we develop Markov Decision Process (MDP) formulation for the considered resource allocation problem and employ diverse Reinforcement Learning (RL) methods for learning optimum decision-making policies adaptive to the IoT environment. Specifically, in this paper we propose an MDP formulation for the considered F-RAN resource allocation problem, and investigate the use of various RL methods, Q-learning (QL), SARSA, Expected SARSA (E-SARSA), and Monte Carlo (MC), for learning the optimal policies of the MDP problem.
We also provide extensive simulation results in various IoT environments of heterogeneous latency requirements to evaluate the performance and adaptivity of the four RL methods.

The remainder of the paper is organized as follows. Section \ref{model} introduces the system model. The proposed MDP formulation for the resource allocation problem is given in Section \ref{problem}. Optimal policies and the related RL algorithms are discussed in Section \ref{policy}. Simulation results are presented in Section \ref{simulation}. Finally, we conclude the paper in Section \ref{conclusion}. A list of notation and abbreviations used throughout the paper is provided in Table \ref{t:notations}.

\vspace{-2mm}
\section{System Model}
\label{model}

\begin{figure}[t]
\centering
\includegraphics[width=.45\textwidth]{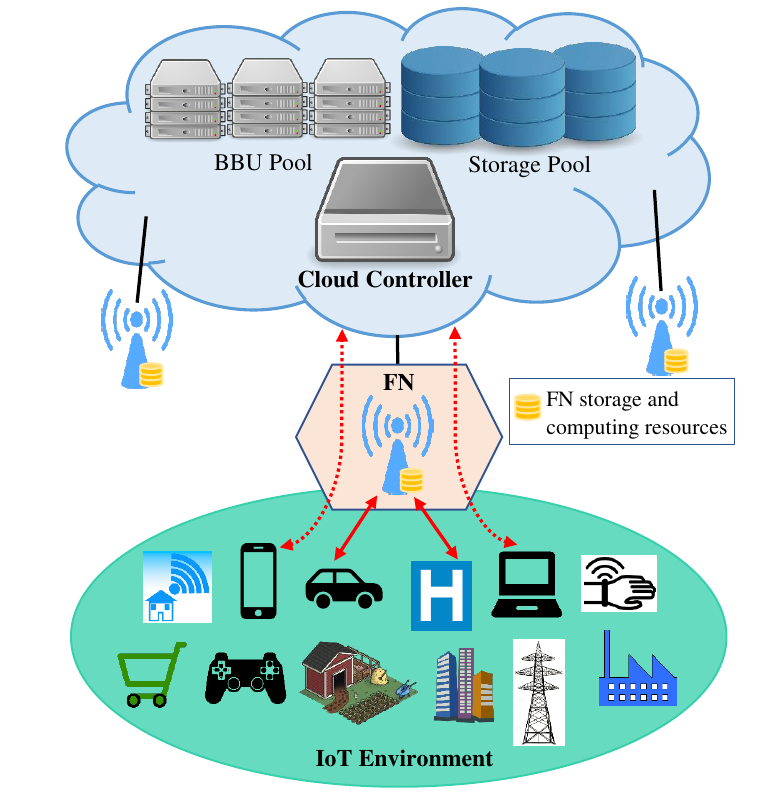}
\vspace{-3mm}
\caption{Fog-RAN system model. The FN serves heterogeneous latency needs in the IoT environment, and is connected to the cloud through the fronthaul links represented by solid lines. Solid red arrows represent local service by FN to satisfy low-latency requirements, and dashed arrows represent referral to the cloud to save limited resources.}
\label{f:model}
\vspace{-4mm}
\end{figure}

We consider the F-RAN structure shown in Fig. \ref{f:model}, in which FNs are connected through the fronthaul to the cloud controller (CC), where a massive computing capability, centralized baseband units (BBUs) and cloud storage pooling are available. To ease the burden on the fronthaul and the cloud, and to overcome the challenge of the increasing number of IoT devices and low-latency applications, FNs are empowered with capability to deliver network functionalities at the edge. Hence, they are equipped with caching capacity, computing and signal processing capabilities. However, these resources are limited, and therefore need to be utilized efficiently. An end user attempts to access the network by sending a request to the nearest FN. The FN takes a decision whether to serve the user locally at the edge using its own computing and processing resources or refer it to the cloud. We consider the FN's computing and processing capacity to be limited to $N$ resource blocks (RBs). User requests arrive sequentially and decisions are taken quickly, so no queuing occurs. 

The QoS requirements of a wireless user are typically given by the latency requirement and throughput requirement. IoT applications have various levels of latency requirement, hence it is sensible for the FN to give higher priority for serving the low-latency applications. To differentiate between similar latency requirements we also consider the risk of failing to satisfy the throughput requirement. This risk is related to the ratio of the achievable throughput to the throughput requirement. The achievable throughput is characterized by the signal-to-noise ratio (SNR) through Shannon channel capacity. Shannon's fundamental limit on the capacity of a communications channel gives an upper bound for the achievable throughput, as a function of available bandwidth ($B$) in Hz and SNR in dB, $C=B+log_2{(1+\text{SNR})}$. 
Hence, we define the utility of an IoT user request to be a function of latency requirement, $l$ (in milliseconds), throughput requirement, $\omega$ (in bits per second), and channel capacity, $C$ (in bits per second), i.e., $u = f(l,\omega,C)$. 
Since the utility should be inversely proportional to the latency requirement, and directly proportional to the achievable throughput ratio, $\mu=C/\omega$, we define utility as
\be
\label{e:utility}
u=\kappa (\mu^\zeta / l^\beta),
\ee
where $\kappa, \zeta, \beta>0$ are mapping parameters. This provides a flexible model for utility. By selecting the parameters $\kappa, \zeta, \beta$ a desired range of $u$ and importance levels for latency and throughput requirements can be obtained. Since F-RAN is intended for satisfying low-latency requirements, typically, more weight should be given to latency by choosing larger $\beta$ values.

FNs should be smart to learn how to decide (serve/refer to the cloud) for each request (i.e., how to allocate its limited resources), so as to achieve the conflicting objectives of maximizing the average total utility of served users over time and minimizing its idle (no-service) time. The system objective can be stated as a constrained optimization problem,
\begin{align}
\begin{split}
\max_{a_0,a_1,\ldots,a_{T-1}} \sum_{t=0}^T \ind{a_t=serve} u_t ~~&\text{and}~~ \min_{a_0,a_1,\ldots,a_{T-1}} \sum_{t=0}^T \ind{a_t=reject} \\ 
\text{subject to}&~~ \sum_{t=0}^T \ind{a_t=serve} = N,
\end{split}
\label{eq:sys_obj}
\end{align}
where $a_t$ denotes the action taken at time $t$ (either serves the request locally or rejects it and refers to cloud), $T$ denotes the termination time when all RBs are filled, $N$ denotes the number of RBs, and $\ind{\cdot}$ is the indicator function taking value $1$ if its argument is true and $0$ if false. The goal is to find the optimum decision policy $\{a_0,a_1,\ldots,a_{T-1}\}$ for an IoT environment which randomly generates $\{u_t\}$. Note that the final decision is always $a_T=serve$ by definition, hence omitted in the policy representation. 

One straightforward approach to deal with this resource allocation problem is to apply a fixed threshold on the user utility. For instance, we can define a threshold rule, such as ``serve  if $ u>5$", if we classify all applications in an IoT environment into ten different utilities $u\in \{1,2,...,10\}$, $10$ being the highest utility. However, such a policy is sub-optimum since the FN will be waiting for a user to satisfy the threshold, which will increase the idle time. The main drawback of this policy is that it cannot adapt to the dynamic IoT environment to achieve the objective. For instance, when the user utilities are almost uniformly distributed, a very selective policy with a high threshold will stay idle most of the time, whereas an impatient policy with a low threshold will in general obtain a low average served utility. A mild policy with threshold 5 may in general perform better than the extreme policies, yet it will not be able adapt to different IoT environments. A better solution for the F-RAN resource allocation problem is to use RL techniques which can continuously learn the environment and adapt the decision rule accordingly.


\vspace{-2mm}
\section{MDP Problem Formulation}
\label{problem}
RL can be thought as the third paradigm of machine learning in addition to the other two paradigms, supervised learning and unsupervised learning. The key point in the proposed RL approach is that FN learns about the IoT environment by interaction and then adapts to it. FN gains rewards from the environment for every action it takes, and once the optimum policy of actions is learned, FN will be able to maximize its expected cumulative rewards, adapt to the IoT environment, and achieve the objective. 

For an access request from a user with utility $u_t$, at time $t$, if the FN decides to take the action $a_t=serve$, which means to serve the user at the edge, then it will gain an immediate reward $r_t$ and one of the RBs will be occupied. Otherwise, for the action $a_t=reject$, which means to reject serving the user at the edge and refer it to the cloud, the FN will maintain its available RBs and get a reward $r_t$. The value of $r_t$ depends on $a_t$ and $u_t$. For tractability, we consider quantized utility values, $u_t \in \{1,2,\ldots,U\}$.  

We define the state $s_t$ of the FN at any time $t$ as
\be
\label{e:state}
s_t=10\,b_t+u_t,
\ee
where $b_t \in \{0, 1, 2,\ldots, N\}$ is the number of occupied RBs at time $t$. Note that the successor state $s_{t+1}$ depends only on the current state $s_t$, the utility $u_{t+1}$ of the next service request, and the action taken ($serve$ or $reject$), satisfying the Markov property $P(s_{t+1}|s_0, ..., s_{t-2}, s_{t-1}, s_t, a_t) = P(s_{t+1}|s_t, a_t)$, i.e., Markov state. Hence, we formulate the Fog-RAN resource allocation problem in the form of a Markov decision process (MDP), which is defined by the tuple $(\cS, \cA, P^a_{ss'}, R^a_{ss'})$, where $\cS$ is the set of all possible states, i.e., $s_t \in \cS$, $\cA$ is the set of actions, i.e., $a_t \in \cA=\{serve, reject\}$, $P^a_{ss'}$ is the transition probability from state $s$ to $s'$ when the action $a$ is taken, i.e., $P^a_{ss'}=P(s'|s,a)$, where $s'$ is a shorthand notation for the successor state, and $R^a_{ss'}$ is the immediate reward received when the action $a$ is taken at state $s$ which ends up in state $s'$, e.g., $r_t = R^{a_t}_{s_t s_{t+1}} \in \cR$. The return $G_t$ is defined as the cumulative discounted rewards received from time $t$ onward and given by
\begin{equation}
\vspace{-1mm}
\label{e:G}
G_t = r_t+\gamma r_{t+1}+\gamma^2 r_{t+2}+...=\sum_{j=0}^{\infty} \gamma^j r_{t+j},
\end{equation}
where $\gamma \in [0,1]$ is the discount factor.
$\gamma$ represents the weight of future rewards with respect to the immediate reward, $\gamma=0$ ignores future rewards, whereas $\gamma=1$ means that future rewards are of the same importance as the immediate rewards. The objective of the MDP problem is to maximize the expected initial return $\Exp[G_0]$.

In the presented MDP, for an FN that has $N$ RBs, there are $U(N+1)$ states, $s_t \in \cS=\{1, 2, 3, \ldots, U(N+1)\}$, where $U$ is the greatest discrete utility level. At the initiation time $t=0$, all RBs are available, i.e., $b=0$, hence from \eqref{e:state}, there are $U$ possible initial states  $s_0 \in\{1, 2,\ldots,U\}$ dependent on $u_0$. The MDP terminates at time $T$ when all RBs are occupied, i.e., $b_T=N$, hence similarly there are $U$ terminal states $s_T \in \{UN+1, UN+2, \ldots, U(N+1)\}$. Note that a policy treating the MDP problem can continue operating after $T$ as in-use RBs become available in time by taking actions similarly to its operation before $T$. 

The reward mechanism $R^a_{ss'}$ is typically chosen by the system designer according to the objective. 
We propose a reward mechanism based on the received utility and the action taken for it. Specifically, at time $t$, based on $u_t$ and $a_t$, the FN receives an immediate reward $r_{t}\in \cR=\{r_{sh}, r_{sl}, r_{rh}, r_{rl}\}$, and moves to the successor state $s_{t+1}$, where $r_{sh}$ is the reward for serving a high-utility request, $r_{sl}$ is the reward for serving a low-utility request, $r_{rh}$ is the reward for rejecting a high-utility request, and $r_{rl}$ is the reward for rejecting a low-utility request. A request is determined as high-utility or low-utility relative to the environment based on a threshold $u_h$, which is a design parameter dependent on the utility distribution in IoT environment. For instance, $u_h$ can be selected as a certain percentile, such as the $\nth{50}$ percentile, i.e., median, of the utilities in the environment. Hence, the proposed reward function is given by
\begin{equation}
r_t = \left\{ 
\begin{array}{ll}
r_{sh} & \text{if}~ a_t=serve, ~u_t \ge u_h \\
r_{rh} & \text{if}~ a_t=reject, ~u_t \ge u_h \\
r_{sl} & \text{if}~ a_t=serve, ~u_t < u_h \\
r_{rl} & \text{if}~ a_t=reject, ~u_t < u_h.
\end{array} 
\right.
\end{equation}

\textbf{\textit{Remark 1:}} Note that the threshold $u_h$ does not have a definitive meaning with respect to the system requirements, i.e., there is no requirement saying that requests with utility lower/greater than $u_h$ must be rejected/served. The goal here is to introduce an internal reward mechanism for the RL approach to facilitate learning the expected future gains, as will be clear later in this section and the following section. For an effective learning performance, the reward mechanism should be simple enough to guide the RL algorithm towards the system objective (see \eqref{eq:sys_obj}) \cite{r32}. That is, its role is not to imitate the system objective closely to make the algorithm achieve it at once, but to resemble it in a simple manner to let the algorithm iteratively achieve a high performance. 

\textbf{\textit{Remark 2:}} Although a threshold $u_h$ is utilized in the proposed reward mechanism, its use is fundamentally different than the straightforward threshold-based policy which always accepts/rejects requests with utility greater/lower than a threshold. While the straightforward threshold-based policy considers only the immediate gain from the current utility, the algorithms tackling the MDP problem, such as the RL algorithms, consider the expected return $\Exp[G_0]$ which includes the immediate reward and expected future rewards. Hence, the threshold $u_h$ does not necessarily cause the algorithm to accept/reject requests with utility greater/lower than $u_h$; it only plays an internal role in learning the expected future rewards.

\begin{table}[t]
\caption{State transitions of 5-RB FN for a sample of IoT requests and random actions with $U=10, u_h=6$}
\label{t:transitions}
\vspace{-4mm}
\begin{center}
\begin{tabular}{ c c c c c c c }
\hline\hline
$t$ &  $u_t$ &  $b_t$ &  $s_t$ &  $a_t$ &  $r_t$ &  $s_{t+1}$\\
\hline
$0$ &  $5$ &  $0$ &  $5$ & $reject$ &  $r_{rl}$ &  $9$\\
$1$ &  $9$ &  $0$ &  $9$ & $serve$ &  $r_{sh}$ &  $13$\\
$2$ &  $3$ &  $1$ &  $13$ & $reject$ &  $r_{rl}$ &  $13$\\
$3$ &  $3$ &  $1$ &  $13$ & $serve$ &  $r_{sl}$ &  $28$\\
$4$ &  $8$ &  $2$ &  $28$ & $serve$ &  $r_{sh}$ &  $36$\\
$5$ &  $6$ &  $3$ &  $36$ & $reject$ &  $r_{rh}$ &  $31$\\
$6$ &  $1$ &  $3$ &  $31$ & $reject$ &  $r_{rl}$ &  $40$\\
$7$ &  $10$ &  $3$ &  $40$ & $serve$ &  $r_{sh}$ &  $47$\\
$8$ &  $7$ &  $4$ &  $47$ & $reject$ &  $r_{rh}$ &  $49$\\
$9$ &  $9$ &  $4$ &  $49$ & $serve$ &  $r_{sh}$ &  $54$\\
$10$ &  $4$ &  $5$ &  $54$ & $\cdot$ &  $\cdot$ &  $\cdot$\\
\hline\hline
\end{tabular}
\end{center}
\end{table}

\begin{figure}[t]
\centering
\includegraphics[width=.5\textwidth]{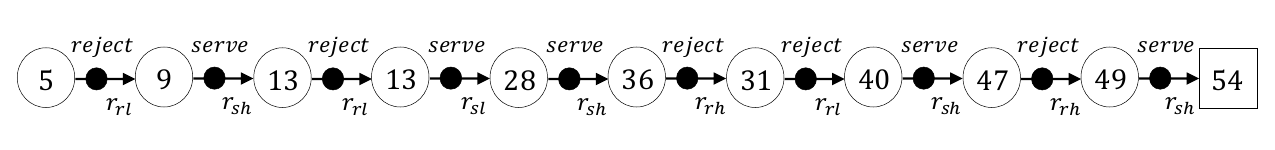}
\caption{State transition graph for the MDP episode given in Table \ref{t:transitions} for an FN with $N=5, U=10, u_h=6$. Non-terminal states and terminal state are represented by circles and squares, respectively, and labeled by the states names. Filled circles represent actions, and arrows show the transitions with corresponding rewards.}
\label{f:episode}
\end{figure}

State transitions for an FN with 5 RBs ($N=5$), $10$ utility levels ($U=10$), and $u_h=6$, a sample of IoT requests with utilities $u_t$, and random actions $a_t$ are shown in Table \ref{t:transitions}. At time $t$, being at state $s_t$, and taking the action $a_t$ will result in getting an immediate reward $r_t$ and moving to the successor state $s_{t+1}$. 
The state transitions in Table \ref{t:transitions} represent an episode of the MDP, it starts at $t=0$ and terminates at $T=10$ with the states $5 \to 9 \to 13 \to 13 \to 28 \to 36 \to 31 \to 40 \to 47 \to 49 \to 54$. The dynamics of this episode is shown through a state transition graph in Fig. \ref{f:episode}, in which non-terminal states and terminal state are represented by circles and squares, respectively, and labeled by the states names, filled circles represent actions, and arrows show the transitions with corresponding rewards.

\vspace{-2mm}
\section{Optimal Policies}
\label{policy}

The state-value function $V(s)$, shown in \eqref{e:V}, represents the long-term value of being in state $s$ in terms of the expected return which can be collected starting from this state onward till termination. Hence, the terminal state has zero value since no reward can be collected from that state, and the value of initial state is equal to the objective function $\Exp[G_0]$. The state value can be viewed also in two parts: the immediate reward from the action taken and the discounted value of the successor state where we move to. Similarly, the action-value function $Q(s,a)$ is the expected return that can be achieved after taking the action $a$ at state $s$, as shown in \eqref{e:Q}. The action value function tells how good it is to take a particular action at a given state. The expressions in \eqref{e:V} and \eqref{e:Q} are known as the Bellman expectation equations for state value and action value, respectively \cite{r32},
\begin{align}
\vspace{-3mm}
\label{e:V}
V(s)&=\Exp[G_t|s_t=s]=\Exp[r_t+ \gamma V(s' )|s], \\
\label{e:Q}
Q(s,a)&=\Exp[G_t|s, a]=\Exp[r_t+ \gamma Q(s', a' )|s,a],
\end{align}
where $a'$ denotes the successor action at the successor state $s'$. 


The objective of the FN in the presented MDP is to utilize the $N$ resource blocks for high-utility IoT applications in a timely manner. This can be done through maximizing the value of initial state, which is equal to the MDP objective $\Exp[G_0]$. To this end, an optimal decision policy is required, which is discussed next.

A policy $\pi$ is a way of selecting actions. It can be defined as the set of probabilities of taking a particular action given the state, i.e.,  $\pi=\{P(a|s)\}$ for all possible state-action pairs. The policy $\pi$ is said to be optimal if it maximizes the value of all states, i.e., $\pi^*=\arg \max\limits_\pi   V_{\pi}(s), \forall s$. Hence, to solve the considered MDP problem, the FN needs to find the optimal policy through finding the optimal state-value function $V^*(s)=\max\limits_\pi V_{\pi}(s)$, which is similar to finding the optimal action-value function $Q^*(s,a)=\max\limits_\pi Q_{\pi}(s,a)$ for all state-action pairs. From \eqref{e:V} and \eqref{e:Q}, we can write the Bellman optimality equations for $V^*(s)$ and $Q^*(s,a)$ as,
\begin{equation}
\label{e:V^*}
V^*(s)= \max\limits_{a\in \cA} Q^*(s,a)=\max\limits_{a \in \cA} \Exp[r_t+\gamma V^*(s') |s,a],
\end{equation}
\begin{equation}
\label{e:Q^*}
Q^*(s,a)= \Exp[r_t+\gamma \max\limits_{a'\in \cA} Q^*(s',a')|s,a].
\end{equation}

\noindent The notion of optimal state-value function $V^*(s)$ greatly simplifies the search for optimal policy. Since the goal of maximizing the expected future rewards is already taken care of the optimal value of the successor state, $V^*(s')$ can be taken out of the expectation in \eqref{e:V^*}. Hence, the optimal policy is given by the best local actions at each state. Dealing with $Q^*(s,a)$ to choose optimal actions is even easier, because with $Q^*(s,a)$ there is no need for the FN to do the one-step-ahead search and instead it picks the best action that maximizes $Q^*(s,a)$ at each state. Optimal actions are defined as follows,
\begin{equation}
\label{e:a^*}
a^*=\arg \max\limits_{a\in \cA} Q^*(s,a)=\arg \max\limits_{a\in \cA} \Exp[r_t|s,a]+\gamma V^*(s'|s,a).
\end{equation}

After discretizing the utility into $U$ levels, the state space becomes tractable with cardinality $|\cS|=U(N+1)$, hence in this case the optimal policy can be learned by estimating the optimal value functions (either \eqref{e:V^*} or \eqref{e:Q^*}) using tabular methods such as model-free RL methods (e.g., Monte Carlo, SARSA, Expected SARSA, and Q-learning), which are also called approximate dynamic programming methods \cite{r32}. Since the expectations involved in value functions are not tractable to find in closed form, we resort to model-free RL methods in this work instead of exact dynamic programming. 
Continuous utility values (see \eqref{e:utility}) would yield infinite dimensional state space, and thus require function approximation methods, such as deep Q-learning \cite{r37}, for predicting the value function at different states, which we leave to a future work. 

In our MDP problem, firstly FN receives a request from an IoT application of utility $u$, then it makes a decision to serve or reject, meaning that the reward for serving $r_s \in \{r_{sh}, r_{sl}\}$ and the reward for rejecting $r_r \in \{r_{rh}, r_{rl}\}$ are known at the time of decision making. Thus, from \eqref{e:V} and \eqref{e:a^*}, the optimal action at state $s$ is given by
\begin{equation}
\label{e:a^*_V}
a^*=
\begin{cases}
\begin{aligned}
serve& \hspace{4pt} \text{if}\  r_s+\gamma \Exp_u[V^*(s'_{serve}=10(b+1)+u_{t+1})] \\ &~~~>r_r+\gamma \Exp_u[V^*(s'_{reject}=10b+u_{t+1})],\\
reject& \hspace{4pt} \text{otherwise},
\end{aligned}
\end{cases} 
\end{equation}

\noindent where $s'_{serve}$ is the successor state when $a=serve$, $s'_{reject}$ is the successor state when $a=reject$, and $\Exp_u$ is the expectation with respect to the utilities $u$ in the IoT environment.

\begin{algorithm}[t]
\caption{Learning Optimum Policy using Monte Carlo}
\label{a:MC}
\begin{algorithmic} [1]
\STATE Select: $\gamma \in [0, 1]$, $\{u_h, r_{sh}, r_{sl}, r_{rh}, r_{rl}\} \in \mathbb{R}$;
\STATE Input: $N$ (number of RBs);
\STATE Initialize: $V(s) \leftarrow 0$, $\forall s$; $\emph{Returns(s)}$ (an array to save states' returns in all iterations);
\FOR{$iteration=0,1,2,...$}
\STATE Initialize: $b \leftarrow 0$;
\STATE Generate an episode: Take actions using \eqref{e:a^*_V} until termination;
\STATE $G(s) \leftarrow$ sum of discounted rewards from $s$ till terminal state for all states appearing in the episode;
\STATE Append $G(s)$ to $\emph{Returns(s)}$;
\STATE $V(s) \leftarrow$ average(Returns(s));
\IF{$V(s)$ converges for all $s$}
\STATE \textbf{break}
\STATE $V^*(s) \leftarrow V(s)$, $\forall s$;
\ENDIF
\ENDFOR
\STATE Use the estimated $V^*(s)$ to find optimal actions using \eqref{e:a^*_V}.
\end{algorithmic}
\end{algorithm}

A popular way to compute the optimal state values, required by the optimal policy as shown in \eqref{e:a^*_V}, is through value iteration by Monte Carlo computations. The procedure to learn the optimal policy from the IoT environment using Monte Carlo is given in Algorithm \ref{a:MC}.
Given the parameters $N$, $\gamma$, $\{u_h, r_{sh}, r_{sl}, r_{rh}, r_{rl}\}$, and the data of IoT users $\{u_t\}$, Algorithm \ref{a:MC} shows how to learn the optimal policy for the considered MDP problem. Note that $\{u_t\}$ can be real data from the IoT environment, as well as from simulations if the probability distribution is known. The \emph{Returns} array at line 2 represents a matrix to save the return of each state at every episode, which corresponds to an iteration. At line 3, we initialize all state values with zeros. Starting from the initial state in each iteration $b=0$, the current state values, which constitutes the current policy, are used to take actions until the terminal state is reached. To promote exploring different states randomized actions can be taken sometimes at line 6 \cite{r32}. $G(s)$ in lines 7 and 8 represents a vector of returns of all states appearing in the episode. Inserting these values into the \emph{Returns} array, the state values are updated by taking the average as shown in line 9. The algorithm stops when all state values converge, the converged values are then used to determine actions as in \eqref{e:a^*_V}.

Similar to \eqref{e:a^*_V}, we can write the optimal action at state $s$ in terms of $Q^*(s,a)$ as follows,
\begin{equation}
\label{e:a^*_Q}
a^*=
\begin{cases}
\begin{aligned}
serve& \hspace{4pt} \text{if}\  Q^*(s,serve)> Q^*(s,reject),\\
reject& \hspace{4pt} \text{otherwise}.
\end{aligned}
\end{cases} 
\end{equation}

\noindent The optimal action-value functions, required by the optimal policy as shown in \eqref{e:a^*_Q}, can be also computed through the value iteration technique using different RL algorithms. The procedure to learn the optimal policy from the IoT environment using the model-free SARSA, E-SARSA, and Q-learning methods is given in Algorithm \ref{a:QL}.

\begin{algorithm}
\caption{Learning Optimum Policy using QL, E-SARSA, and SARSA }
\label{a:QL}
\begin{algorithmic} [1]
\STATE Select: \{$\gamma, \epsilon\} \in [0, 1]$, $\alpha \in (0, 1]$, $n \in \{1,2,...\}$;
\STATE Input: $N$ (number of RBs);
\STATE Initialize: $Q(s,a)$ arbitrarily in $\mathbb{Q}$, $\forall (s, a)$;
\STATE Initialize: $b \leftarrow 0$;
\FOR{$t=0,1,2,...$}
\STATE Take action $a_t$ according to $\pi$ (e.g., $\epsilon$-greedy), and store $r_t$ and $s_{t+1}$;
\IF{$t \ge n-1$}
\STATE $\tau \leftarrow t+1-n$;
\STATE QL: $G \leftarrow \sum_{j=\tau}^{t+1} \gamma^{\,(j-\tau)} r_{j} +\gamma^{\,n}\max\limits_{a} Q(s_{t+1},a)$;
\STATE E-SARSA: $G \leftarrow \sum_{j=\tau}^{t+1} \gamma^{\,(j-\tau)} r_{j} +\gamma^{\,n} \Exp_a[Q(s_{t+1},a)]$;
\STATE SARSA: $G \leftarrow \sum_{j=\tau}^{t+1} \gamma^{\,(j-\tau)} r_{j} +\gamma^{\,n}Q(s_{t+1},a_{t+1})$;

\STATE $Q(s_\tau,a_\tau) \leftarrow Q(s_\tau,a_\tau)+\alpha[G-Q(s_\tau,a_\tau)]$;
\STATE Update $\mathbb{Q}$ with $Q(s_\tau,a_\tau)$;
\ENDIF
\IF{$Q(s,a)$ converges for all $(s,a)$}
\STATE $Q^*(s,a) \leftarrow Q(s,a)$;
\STATE \textbf{break}
\ENDIF
\ENDFOR
\STATE Use $Q^*(s,a)$ estimated in $\mathbb{Q}$ for $\pi^*$ using \eqref{e:a^*_Q}
\end{algorithmic}
\end{algorithm}

\indent Algorithm \ref{a:QL} shows how FN learns the optimal policy for the MDP by estimating $Q^*(s,a)$ using QL, E-SARSA, and SARSA methods. The step size parameter $\alpha$ represents the weight we give to the change in our experience, i.e., the learning rate, $\epsilon$ is the probability of making a random action for exploration, and the batch size $n$ represents the number of time steps after which we update the $Q(s,a)$ values. The $\mathbb{Q}$ array at line 3 represents a matrix to save the updated values of the action-value functions of all states and actions in each iteration. 
In each iteration, we take an action, observe and store the collected reward and the successor state. Actions are taken according to a policy $\pi$ such as the $\epsilon$-greedy policy in line 6, in which a random action with probability $\epsilon$ is taken to explore new rewards, and an optimal action (see \eqref{e:a^*_Q}) is taken with probability $(1-\epsilon)$ to maximize the rewards; with $\epsilon=0$, the policy becomes greedy. The condition at line 7 represents the time, in terms of the batch size, at which we start updating the $Q$ values of the actions taken in the previously visited states. The way target $G$ is computed for QL, E-SARSA and SARSA is shown at lines 9-11. $G$ represents the return collected starting from time $(t+1-n)$ to $n$ time-steps ahead, and it contains two parts, the discounted collected rewards and a function of the action-value for future rewards. The latter part changes for QL, E-SARSA and SARSA. For QL, the maximum action-value is used considering all possible actions which can be taken from the state at $t+1$. Whereas, E-SARSA uses the expected value of $Q(s_{t+1},a)$ over possible actions at state $s_{t+1}$, and SARSA uses $Q(s_{t+1},a_{t+1})$ considering the action that will be taken at time $t+1$ according to the current policy. The way to update the action-value is shown at line 12, where $\tau$ is the time whose $Q$ estimate is being updated. At line 13, the matrix $\mathbb{Q}$ is updated with the new $Q$ value and used to make future decisions. The algorithm stops when all $Q$ values converge. The converged values represent the optimal action values $Q^*$ which are then used to determine optimal actions as in \eqref{e:a^*_Q}.

\section{Simulations}
\label{simulation}
We next provide simulation results to evaluate the performance of FN when implementing the RL methods, Q-learning, SARSA, Expected-SARSA, and Monte Carlo, given in Algorithms \ref{a:MC} and \ref{a:QL}. We also compare the RL-based FN performance with the FN performance when a fixed thresholding algorithm is employed. 
We evaluate the performances in various IoT environments with different compositions of IoT latency requirements.
For brevity, we do not consider the effect of ratio of the achievable throughput to the throughput requirement in assessing the utility of a service request. Specifically, we consider 10 utility classes with different latency requirements to exemplify the variety of IoT applications in an F-RAN setting. That is, we consider $\zeta=0, \beta=1, \kappa=1$ in \eqref{e:utility}, and discretize the latency-based utility to 10 classes ($U=10$). The utility values $1, 2, ..., 10$ may represent the following IoT applications, respectively: smart farming, smart retail, smart home, wearables, entertainment, smart grid, smart city, industrial Internet, autonomous vehicles, and connected health.
By changing the composition of utility classes, we generate 19 scenarios of IoT environments, 6 of which are summarized in Table \ref{t:environments}. Higher density of high-utility users makes the IoT environment richer in terms of low-latency IoT applications.

\begin{table}
\caption{Utility distributions for various IoT environments with heterogeneous latency requirements}
\label{t:environments}
\vspace{-4mm}
\begin{center}
\begin{tabular}{ c c c c c c c }
\hline\hline
$$ & $\mathcal{E}_1$ & $\mathcal{E}_4$ & $\mathcal{E}_7$ & $\mathcal{E}_{10}$ & $\mathcal{E}_{15}$ & $\mathcal{E}_{19}$\\
\hline
$P(u=1)$ &  $0.015$ &  $0.012$ &  $0.01$ &  $0.008$ &  $0.004$ &  $0.001$\\
$P(u=2)$ &  $0.073$ &  $0.062$ &  $0.05$ &  $0.038$ &  $0.019$ &  $0.004$\\
$P(u=3)$ &  $0.365$ &  $0.308$ &  $0.25$ &  $0.192$ &  $0.096$ &  $0.019$\\
$P(u=4)$ &  $0.292$ &  $0.246$ &  $0.2$ &  $0.154$ &  $0.077$ &  $0.015$\\
$P(u=5)$ &  $0.205$ &  $0.172$ &  $0.14$ &  $0.108$ &  $0.054$ &  $0.011$\\
$P(u=6)$ &  $0.014$ &  $0.057$ &  $0.1$ &  $0.142$ &  $0.214$ &  $0.271$\\
$P(u=7)$ &  $0.013$ &  $0.051$ &  $0.09$ &  $0.129$ &  $0.193$ &  $0.244$\\
$P(u=8)$ &  $0.011$ &  $0.046$ &  $0.08$ &  $0.114$ &  $0.171$ &  $0.217$\\
$P(u=9)$ &  $0.009$ &  $0.034$ &  $0.06$ &  $0.086$ &  $0.129$ &  $0.163$\\
$P(u=10)$ &  $0.003$ &  $0.012$ &  $0.02$ &  $0.029$ &  $0.043$ &  $0.055$\\
$\rho=P(u>5)$ &  $5\%$ &  $20\%$ &  $35\%$ &  $50\%$ &  $75\%$ & $95\%$\\
$\bar{u}$ &  $3.82$ &  $4.4$ &  $4.97$ &  $5.55$ &  $6.5$ & $7.27$\\
\hline\hline
\vspace{-10mm}
\end{tabular}
\end{center}
\end{table}

Denoting an IoT environment of a particular utility distribution with $\mathcal{E}$, we show in Table \ref{t:environments} the statistics of $\mathcal{E}_1$, $\mathcal{E}_4$, $\mathcal{E}_7$, $\mathcal{E}_{10}$, $\mathcal{E}_{15}$, and $\mathcal{E}_{19}$. The first $10$ rows in the table provide detailed information about the proportion of each utility class in an IoT environment corresponding to a latency requirement. The last two rows illustrate the quality or richness of IoT environments, where $\rho$ is the probability of a utility being greater than $5$, and $\bar{u}$ is the mean value of utilities in the environment. In the considered 19 scenarios, $\rho$ increases by 0.05 from 5\% to 95\% for $\mathcal{E}_1, \mathcal{E}_2, ..., \mathcal{E}_{19}$ respectively. The remaining 13 scenarios have statistics proportional to their $\rho$ values. We started with a general scenario given by $\mathcal{E}_7$, and changed $\rho$ to obtain the other scenarios.

The simulation parameters shown in Table \ref{t:sim} are used for the presented results in this section. The rewards $\cR=\{r_{sh}, r_{sl}, r_{rh}, r_{rl}\}$ are chosen to facilitate learning the optimal policy. We consider that the FN is equipped with computing, signal processing and storage resources of $15$ resource blocks (RBs), i.e., $N=15$. In a particular environment $\mathcal{E}$, the threshold that defines ``high utility" is set to the mean of all utilities, i.e., $u_h=\bar{u}$. We applied the greedy policy in our simulations, hence $\epsilon=0$.

\begin{table}
\caption{Summary of simulation parameters and their values}
\label{t:sim}
\vspace{-5mm}
\begin{center}
\begin{tabular}{c c c}
\hline\hline
Parameter &Description & Value\\
\hline
$\gamma$  &discount factor &  $0.7$\\
$\alpha$    &learning rate &  $0.01$\\
$\epsilon$ &probability of random action &  $0$\\
$\theta$ &penalty of idle time &  $1$\\
$n$ &  batch/step size &  $1$\\
$N$ &  total number of resource blocks of FN &  $15$\\
$r_{sh}$ &  reward for serving high-utility user &  $2$\\
$r_{sl}$ &  reward for serving low-utility user &  $-1$\\
$r_{rh}$ &  reward for rejecting high-utility user &  $-2$\\
$r_{rl}$ &  reward for rejecting low-utility user &  $1$\\
$u_h$ &  the threshold for ``high-utility" &  mean\\
\hline\hline
\end{tabular}
\end{center}
\end{table}

\begin{figure}[t]
\centering
\includegraphics[width=.5\textwidth]{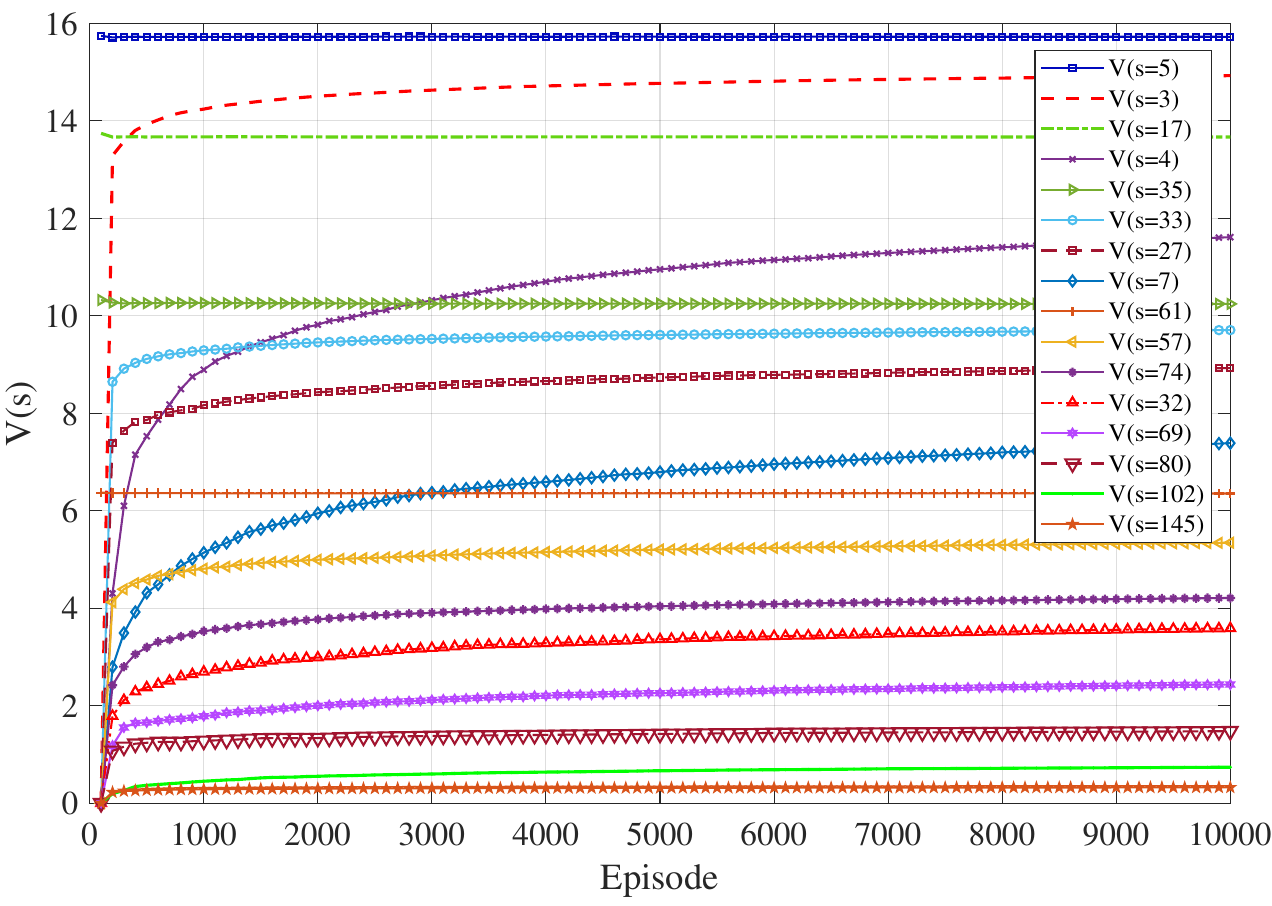}
\caption{Learning optimum policy of the MDP by applying the Monte Carlo method given by Algorithm \ref{a:MC} to obtain the optimal state values required in \eqref{e:a^*_V}. The IoT environment $\mathcal{E}_7$ is considered, and the FN is equipped with 15 RBs. The 16 state values shown in the figure are a sample of the 150 non-terminal state values.}
\label{f:V}
\vspace{-6mm}
\end{figure}

We firstly consider the MDP formulation for the IoT environment given by scenario $\mathcal{E}_7$ shown in Table \ref{t:environments}. By interaction with the environment, the FN updates the state value functions which converge to the optimum policy. Fig. \ref{f:V}, shows how the FN learns the optimal policy using the Monte Carlo (MC) method given in Algorithm \ref{a:MC} to estimate the optimal state values. With 15 RBs, there are 160 states, the last 10 of which are terminal states with $b=15$ for which $V(s)=0$. The state-value functions of 16 states are given in \ref{f:V}. The remaining states have values within a standard deviation $\sigma=0.5$ of the selected 16 states. It is seen that for most of the states the state values converges the optimal value $V^*(s)$ after about 5000 iterations. 
This number can be easily exceeded by the number of requests received by FN during a busy hour from a variety of IoT applications \cite{r1}.

\begin{figure}[t]
\centering
\includegraphics[width=.5\textwidth]{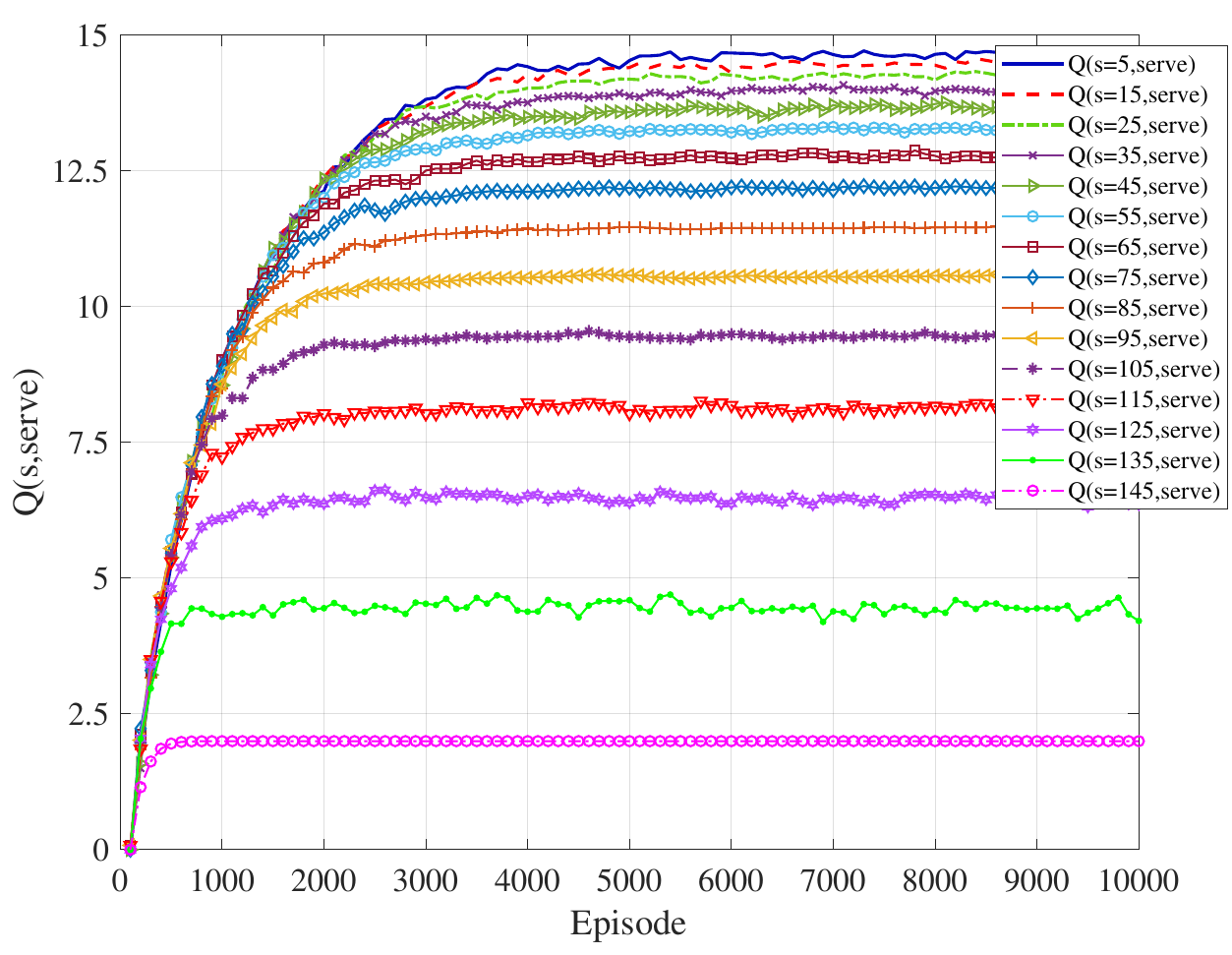}
\caption{Learning the optimal action-value function $Q^*(s,serve)$ required in \eqref{e:a^*_Q} using the Q-learning method given by Algorithm \ref{a:QL}. Q-values converge to the optimal values after around $4000$ episodes. The IoT environment $\mathcal{E}_7$ is considered, and the FN is equipped with 15 RBs.}
\label{f:Qserve}
\vspace{-4mm}
\end{figure}

\begin{figure}[t]
\centering
\includegraphics[width=.5\textwidth]{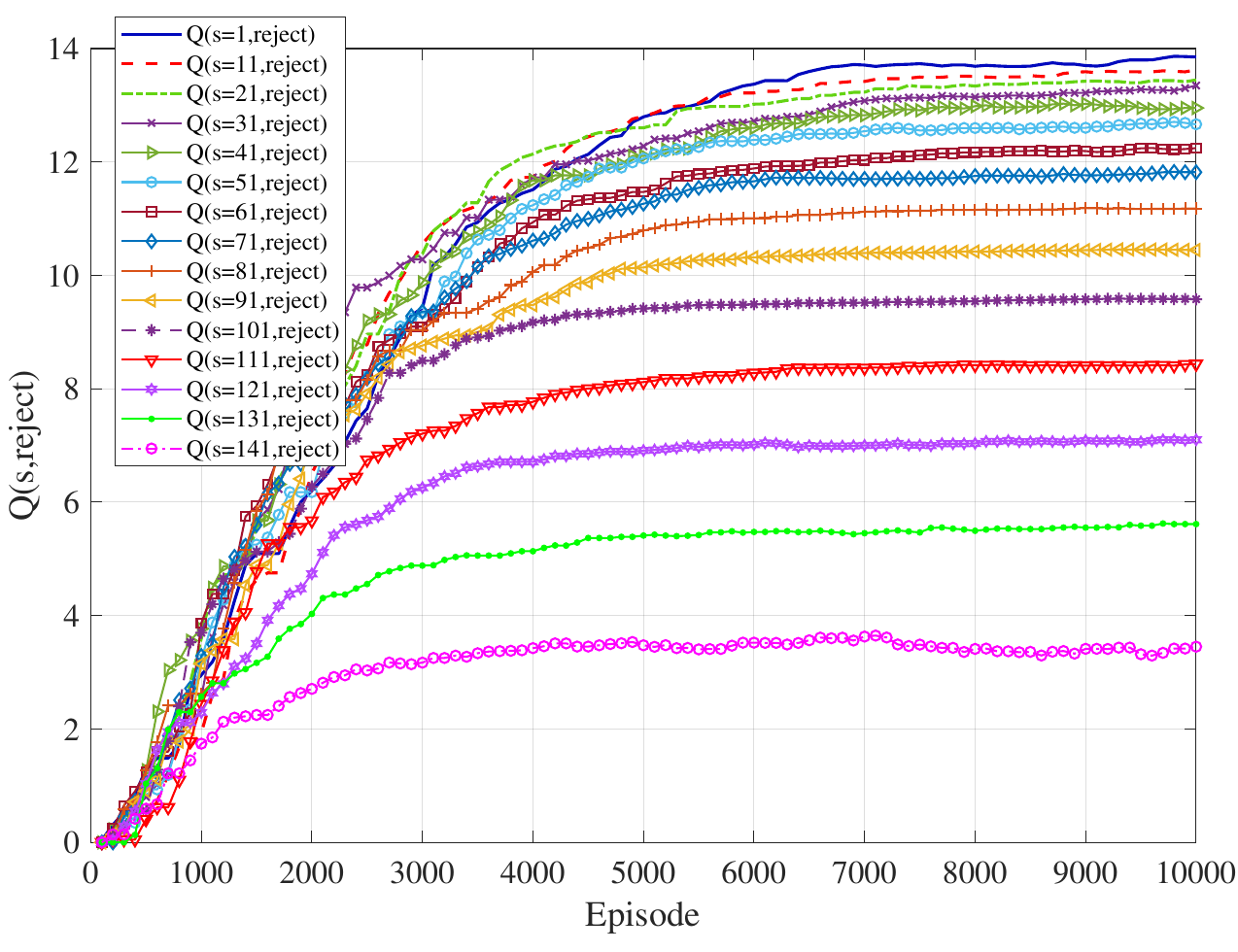}
\caption{Learning the optimal action-value function $Q^*(s,reject)$ required in \eqref{e:a^*_Q} using the Q-learning method given by Algorithm \ref{a:QL}. Q-values converge to the optimal values after around $5000$ episodes. The IoT environment $\mathcal{E}_7$ is considered, and the FN is equipped with 15 RBs.}
\label{f:Qreject}
\vspace{-4mm}
\end{figure}

We next apply SARSA, Expected SARSA and QL in the IoT environment $\mathcal{E}_7$, for learning the optimal policy in \eqref{e:a^*_Q} using the estimated $Q^*(s,a)$ in Algorithm \ref{a:QL}. The convergence of $Q(s,serve)$ and $Q(s,reject)$ when using QL is shown in Figs. \ref{f:Qserve} and \ref{f:Qreject}, respectively. In our MDP problem, QL converges slightly faster than E-SARSA, SARSA and MC since it implements a greedy approach by selecting the maximum $Q(s',a')$ when updating the return $G_t$ as shown in Algorithm \ref{a:QL}. However, this is not a general rule as it depends on the nature of each problem. There are many factors affecting the convergence rate, e.g., large values of the learning rate $\alpha$ make the Q-values bounce around a mean value, whereas small values causes it to converge slowly. 
Unnecessary exploration makes the convergence slower, controlled by the $\epsilon$ value in the $\epsilon$-greedy policy. The step size $n$ after which we update the the state values or Q-values affects also the convergence dependent on the problem. For instance, MC updates the state values at the end of an episode regardless of how long it is, which makes it slower to exploit the updated state values in making better actions, whereas QL, SARSA and E-SARSA using $n=1$ update the Q-value every time step. Unlike MC, the FN needs to keep updating two Q-values for each state instead of updating one state value. Hence, we have 300 Q-values to update in order to learn the optimal policy. 

Recall that the FN objective is to maximize the expected total served utility and minimize the expected termination time, as shown in \eqref{eq:sys_obj}. Hence, to compare the performance of FN when using QL, SARSA, E-SARSA and MC provided in Algorithms \ref{a:MC} and \ref{a:QL} with the performance of a fixed-threshold algorithm, which does not learn from the interactions with environment, we define an objective performance metric $R$ as

\begin{equation}
\label{e:R}
R=\Exp\left[\sum_{m=1}^{M}u_m-\theta(T-M)\right],
\end{equation}

\noindent where a served utility is denoted with $u_m$, the number of served IoT requests in an episode is denoted with $M$, $(T-M)$ represents the total idle time for RBs, and $\theta$ is a penalty for being idle, selected as $1$ in the following comparisons.

\begin{figure}[t]
\centering
\includegraphics[width=.5\textwidth]{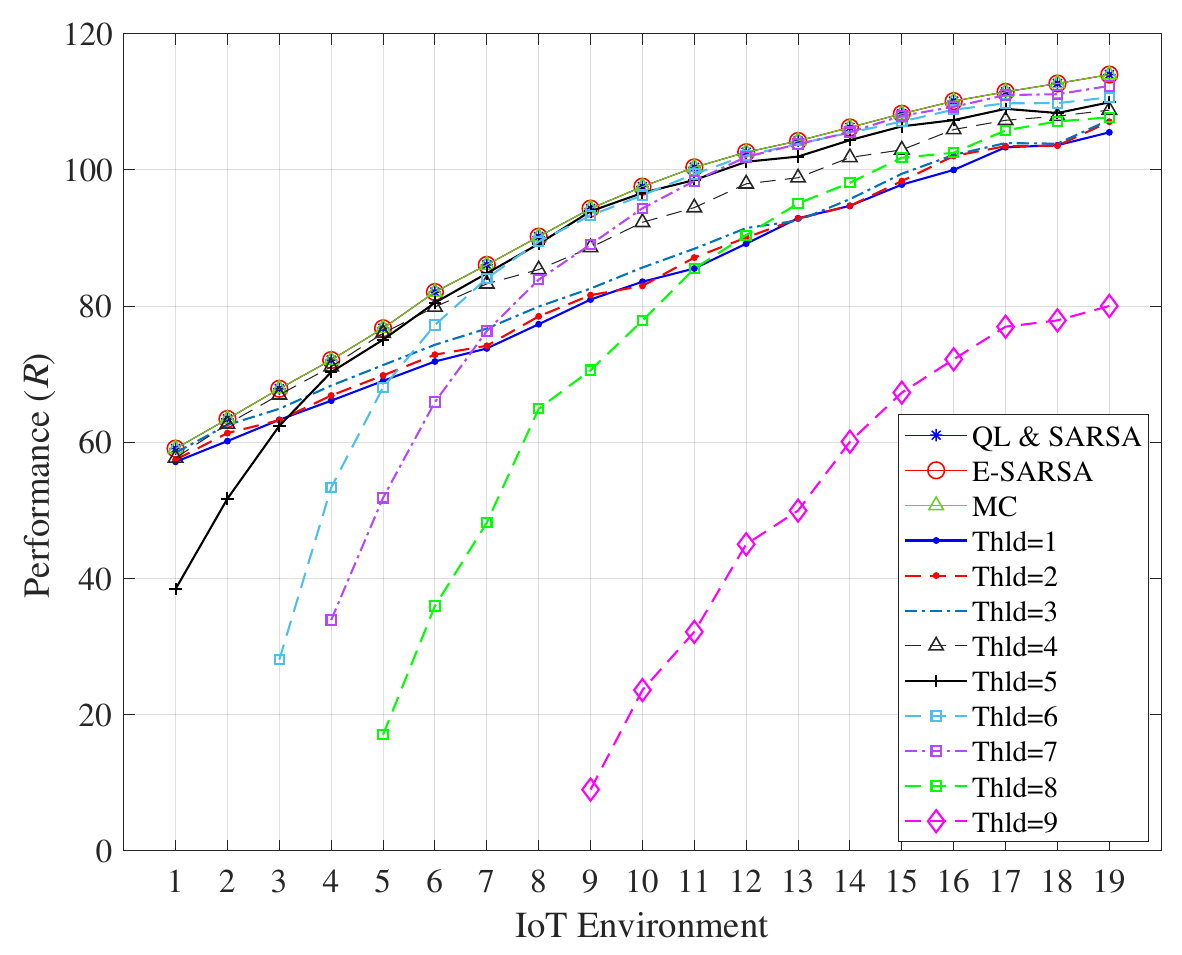}
\caption{The performance in terms of $R$ for the FN with $N=15$, in various IoT environments when applying the RL methods (QL, SARSA, E-SARSA and MC) given in Algorithms \ref{a:MC} and \ref{a:QL}, and the fixed-threshold algorithm with different thresholds. RL methods' performances are indistinguishable here, and better than the fixed thresholds in all environments thanks to their learning/adaptation capability.}
\label{f:R}
\vspace{-4mm}
\end{figure}

\begin{figure}[t]
\centering
\includegraphics[width=.5\textwidth]{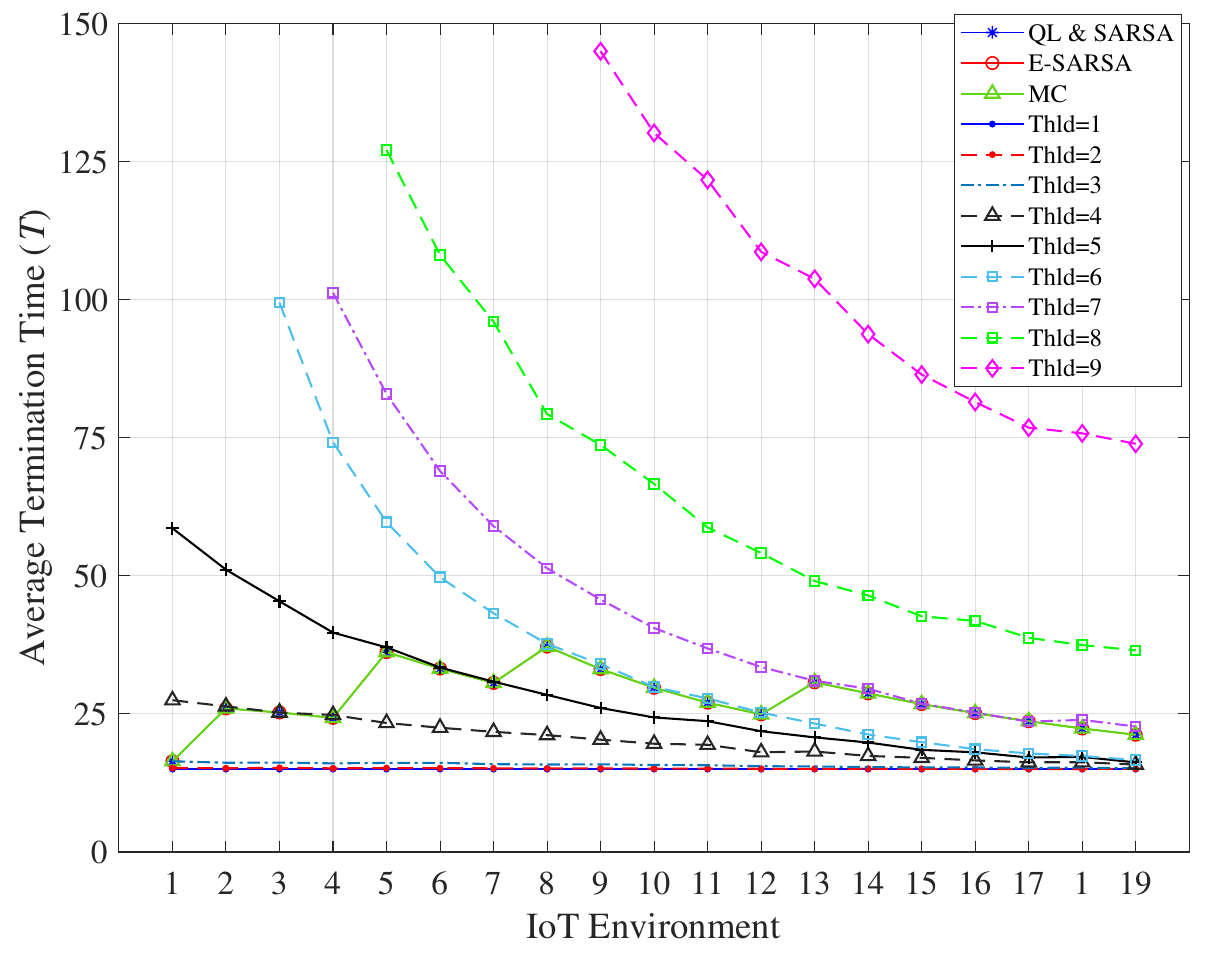}
\caption{The average termination time $T$ for FN with $N=15$ in various IoT environments when applying the RL methods (QL, SARSA, E-SARSA and MC) given by Algorithms \ref{a:MC} and \ref{a:QL}, and the fixed-threshold algorithm with different thresholds. RL methods manage to have a steady termination time in all environments.}
\label{f:T}
\vspace{-4mm}
\end{figure}

We compare the performance of the RL methods, in terms of $R$, with that of the fixed-threshold algorithm in the 19 IoT environments. The fixed threshold-based algorithm uses the same threshold regardless of the environment. For the RL methods, we consider the simulation setup shown in Table \ref{t:sim}, and for the fixed-threshold algorithm we consider all possible thresholds $1, 2, ..., 10$.
As shown in Figs. \ref{f:R} and \ref{f:T}, the RL methods exhibit the best performance as they learn how to balance early termination with higher total served utilities. It never terminates too early or too late ($T\approx27$ for all environments as seen in Fig. \ref{f:T}), as opposed to the fixed-threshold algorithm which is not adaptive to the environment. As seen in Fig. \ref{f:R}, the performance of fixed-threshold algorithm with thresholds $1, 2, 3, 8, 9$ are steadily below that of the RL algorithms. The average termination time for thresholds 1, 2, and 3 is about 15 which is the minimum termination time, though they could not achieve good performance. Threshold 4 has a comparable performance to RL for the environments $\mathcal{E}_2-\mathcal{E}_5$, after which its performance starts to decline. Although thresholds $5, 6, 7$ have good performances close to RL for environments with medium to high $\rho$, they perform far from RL for IoT environments with small $\rho$. The performance of threshold 10 is much worse than threshold 9 for all environments due to the long termination time which exceeds 280, thus it does not appear in Figs. \ref{f:R} and \ref{f:T}.

\begin{figure}[t]
\centering
\includegraphics[width=.5\textwidth]{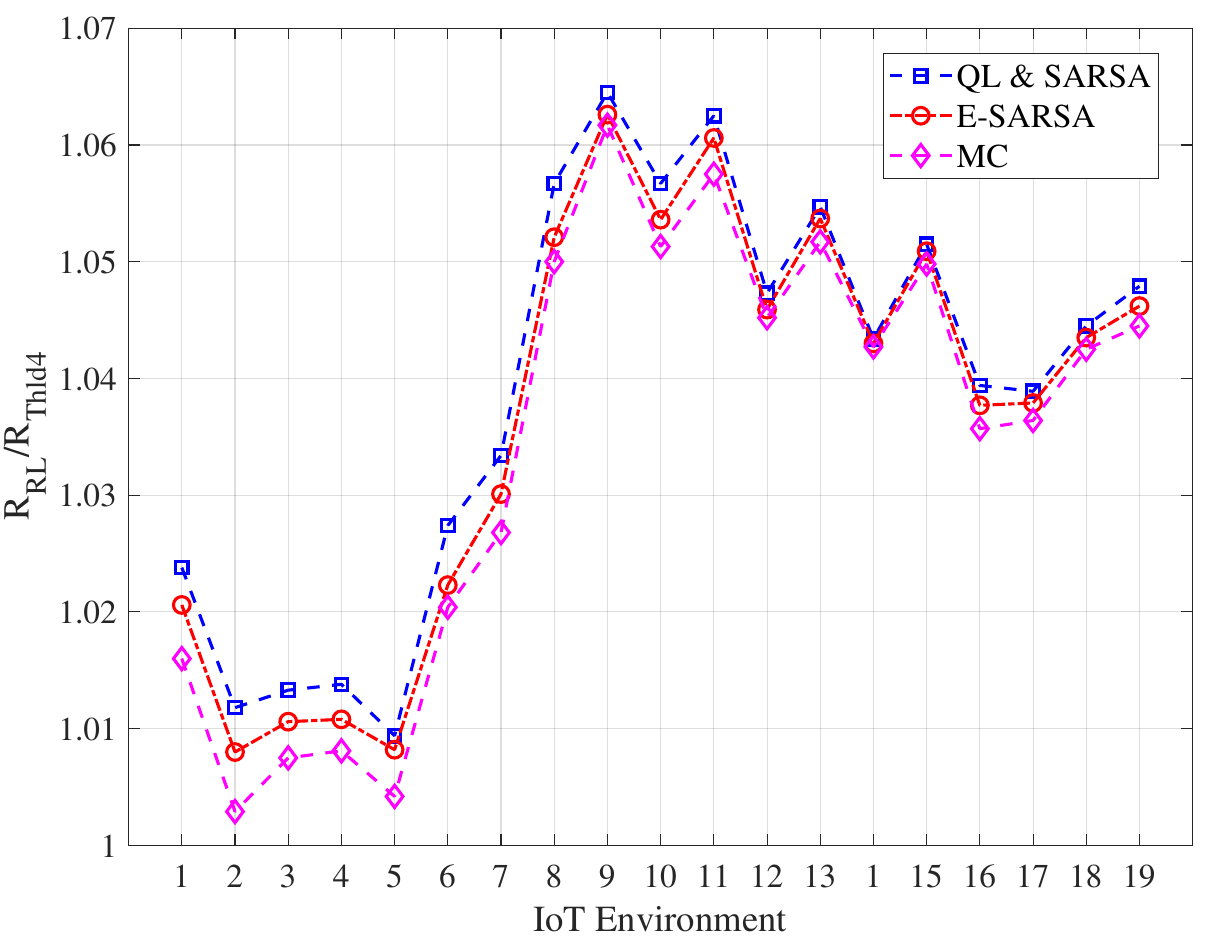}
\caption{Comparison between the performance of RL methods in terms of relative performance with respect to the fixed-threshold algorithm with threshold 4. QL and SARSA coincide due to the greedy policy used in the simulations.}
\label{f:ratio}
\vspace{-4mm}
\end{figure}

The performance of the RL methods is very close to each other, hence it is not easy to distinguish them in Figs. \ref{f:R} and \ref{f:T}. For a clearer view, Fig. \ref{f:ratio} compares the performance of the four RL methods in terms of the performance ratio with respect to performance of threshold 4. QL has the best performance with an average performance ratio of 104\% in all IoT environments with a peak of 106\% in $\mathcal{E}_9$, followed by E-SARSA and MC. SARSA has the same performance as QL because greedy policy, i.e., $\epsilon=0$, was used.

\section{Conclusions}
\label{conclusion}
We proposed a Markov Decision Process (MDP) formulation for the resource allocation problem in Fog RAN for IoT services with heterogeneous latency requirements. Several reinforcement learning (RL) methods, namely Q-learning, SARSA, Expected SARSA, and Monte Carlo, were discussed for learning the optimum decision-making policy adaptive to the IoT environment. Their superior performance over conventional fixed-threshold methods, and adaptivity to the IoT environment were verified through extensive simulations. The RL methods strike a right balance between the two conflicting objectives, maximize the average total served utility vs. minimize the fog node'€™s idle time, which helps utilize fog node's limited resource blocks efficiently. As future work we consider expanding the presented resource allocation framework to more challenging scenarios such as dynamic resource allocation with heterogeneous service times and number of resource blocks needed, and collaborative resource allocation with multiple fog nodes.

\bibliographystyle{IEEEtran}
\bibliography{refs}


\onecolumn
\begin{table}
\caption{Summary of notations and abbreviations}
\label{t:notations}
\begin{center}
\begin{tabular}{ c l c l }
\hline\hline
Notation &Description &Notation &Description\\
\hline
IoT &Internet of things &$a_t$  &action taken at time $t$\\
F-RAN &fog radio access network &$a'$  &action from successor state\\
C-RAN &cloud radio access network &$a^*$  &optimal action\\
5G &fifth generation &$r_t$  &reward received at time $t$\\
URLLC &ultra-reliable low-latency communication &$r_{s}$ &reward for serving\\
FN &fog node &$r_{sh}$ &reward for serving high-utility user\\
SNR &signal-to-noise ratio &$r_{sl}$ &reward for serving low-utility user\\
MDP &Markov decision process &$r_{r}$ &reward for rejecting\\
RL &reinforcement learning &$r_{rh}$ &reward for rejecting high-utility user\\
ML &machine learning &$r_{rl}$ &reward for rejecting low-utility user\\
QL &Q-learning &$\pi$  &policy for taking actions\\
E-SARSA &Expected SARSA &$\pi^*$  &optimal policy\\
MC &Monte Carlo &$G_t$ &return from time $t$ onward\\
RF &radio frequency &$V(s)$  &state-value function of $s$\\
MIMO &multi input multi output &$V_\pi(s)$  &state-value function following policy $\pi$\\
CC &cloud controller &$V^*(s)$  &optimal state-value function\\
RRU &remote radio unit &$Q(s,a)$  &action-value function\\
BBU &baseband unit &$Q^*(s,a)$  &optimal action-value function\\
QoS &quality of service &$Q_\pi(s,a)$  &action-value function following policy $\pi$\\
MBB &mobile broadband &$T$ &termination time\\
RB &resource block &$P^a_{ss'}$ &transition probability to $s'$ given $s,a$\\
Thld &threshold &$R^a_{ss'}$ &reward received for taking $a$ from $s$ given $s'$\\
$\cS$ &set of states &$N$ &total number of FN's resource blocks\\
$\zeta, \beta, \kappa$  &parameters for utility computation &$\Exp_u$  &Expectation with respect to $u$\\

$\cA$ &set of actions &$b_t$ &number of occupied RBs at time $t$\\
$\theta$ &penalty for staying idle&$\gamma$  &discount factor\\
$M$  &number of served IoT requests in an episode&$\alpha$    &learning rate\\

$C$  &channel capacity in bps &$\epsilon$ &probability of random action\\
$B$  &frequency bandwidth in Hz &$n$ &  batch size\\
$l$  &latency in milliseconds &$\mathbb{Q}$  &an array for updated $Q(s,a)$, for all $(s,a)$\\
$\omega$  &throughput requirement in bps &$\tau$  &time whose $Q$ estimate is being updated\\
$u$  &user utility &$\mathcal{E}$  &IoT environment\\
$u_h$  &threshold for defining ``high utility" &$\bar{u}$  &mean value of utilities in Iot environment\\
$\mu$  &achievable throughput ratio &$\sigma$  &standard deviation\\
$\rho$  &probability of $u>5$ in Iot environment &$R$  &objective performance metric\\
$s_t$  &state at time $t$ &$u_m$  &utility of served IoT request\\
$s'$  &successor state & \\
\hline\hline
\end{tabular}
\end{center}
\end{table}

\end{document}